# Efficient and Universal Merkle Tree Inclusion Proofs via OR Aggregation

Oleksandr Kuznetsov [1,2*], Alex Rusnak [1], Anton Yezhov [1], Dzianis Kanonik [1], Kateryna Kuznetsova [1], Oleksandr Domin [1]

[1] Proxima Labs, 1501 Larkin Street, suite 300, San Francisco, USA
[2] Department of Political Sciences, Communication and International Relations, University of Macerata, Via Crescimbeni, 30/32, 62100 Macerata, Italy

*Corresponding author: Oleksandr Kuznetsov
e-mails: kuznetsov@karazin.ua, kuznetsov@proxima.one

Contributing authors:

Alex Rusnak
e-mail: alex@proxima.one
https://orcid.org/0009-0000-1586-8003

Anton Yezhov
e-mail: anton@proxima.one
https://orcid.org/0009-0004-6380-5233

Dzianis Kanonik
e-mail: denis@proxima.one
https://orcid.org/0009-0009-6102-3736

Kateryna Kuznetsova
e-mail: kateryna@proxima.one
https://orcid.org/0000-0002-5605-9293

Oleksandr Domin
e-mail: dyomin@proxima.one
https://orcid.org/0009-0009-1591-137X

**Abstract**: Zero-knowledge proofs have emerged as a powerful tool for enhancing privacy and security in blockchain applications. However, the efficiency and scalability of proof systems remain a significant challenge, particularly in the context of Merkle tree inclusion proofs. Traditional proof aggregation techniques based on AND logic suffer from high verification complexity and data communication overhead, limiting their practicality for large-scale applications. In this paper, we propose a novel proof aggregation approach based on OR logic, which enables the generation of compact and universally verifiable proofs for Merkle tree inclusion. By aggregating proofs using OR logic, we achieve a proof size that is independent of the number of leaves in the tree, and verification can be performed using any single valid leaf hash. This represents a significant improvement over AND aggregation, which requires the verifier to process all leaf hashes. We formally define the OR aggregation logic, describe the process of generating universal proofs, and provide a comparative analysis demonstrating the advantages of our approach in terms of proof size, verification data, and universality. Furthermore, we discuss the potential of combining OR and AND aggregation logics to create complex acceptance functions, enabling the development of expressive and efficient proof systems for various blockchain applications. The proposed techniques have the potential to significantly enhance the scalability, efficiency, and flexibility of zero-knowledge proof systems, paving the way for more practical and adaptive solutions in the blockchain ecosystem.

## 1. Introduction

Zero-knowledge proofs (ZKPs) have garnered significant attention in recent years due to their ability to enhance privacy and security in various applications, particularly in the domain of blockchain technology [1, 2]. ZKPs allow one party (the prover) to convince another party (the verifier) that a statement is true without revealing any additional information beyond the validity of the statement itself [3]. This property makes ZKPs a powerful tool for enabling secure and privacy-preserving transactions, smart contracts, and other applications in blockchain systems [4–7].

One of the fundamental building blocks of many blockchain protocols is the Merkle tree [8–10], a data structure that enables efficient and secure verification of large datasets. Merkle trees are used to store transactions, account balances, and other critical information in a compact and tamper-evident manner [8]. To prove the inclusion of a specific data element within a Merkle tree, a prover must provide a Merkle proof, which consists of a path of hashes from the leaf node (representing the data element) to the root of the tree [8–10].

However, as blockchain networks scale and the number of data elements grows, the size and complexity of Merkle proofs can become a bottleneck in terms of verification efficiency and data communication overhead [11–13]. Traditional proof aggregation techniques based on AND logic, where multiple proofs are combined into a single proof, have been proposed to address this issue [14, 15]. Nevertheless, AND aggregation still requires the verifier to process all the leaf hashes, leading to high verification complexity, especially for large Merkle trees.

In this paper, we propose a novel proof aggregation approach based on OR logic, which enables the generation of compact and universally verifiable proofs for Merkle tree inclusion. Our scheme leverages the properties of OR logic to create a single, aggregated proof that can be verified using any valid leaf hash, thus significantly reducing the verification complexity and data communication overhead compared to AND aggregation. We formally define the OR aggregation logic, describe the process of generating universal proofs, and provide a comparative analysis demonstrating the advantages of our approach in terms of proof size, verification data, and universality.

Furthermore, we explore the potential of combining OR and AND aggregation logics to create complex acceptance functions at the proof generation level. This flexibility enables the development of expressive and efficient proof systems that can cater to various business logic requirements in blockchain applications. By leveraging the strengths of both OR and AND aggregation, our approach opens up new possibilities for constructing scalable, flexible, and privacy-preserving solutions in the blockchain ecosystem.

The rest of this paper is organized as follows: Section 2 provides the necessary background on zero-knowledge proofs, Merkle trees, and proof aggregation techniques. Section 3 introduces our proposed OR aggregation scheme, including the formal definitions and the process of generating universal proofs. Section 4 presents a comparative analysis of our approach with traditional AND aggregation and other related works, discussing the potential applications and extensions of our scheme, particularly in the context of complex acceptance functions. Finally, Section 5 concludes the paper and outlines future research directions.

## 2. Background

### 2.1 Zero-Knowledge Proofs

Zero-knowledge proofs (ZKPs) are cryptographic protocols that allow a prover to convince a verifier that a statement is true without revealing any additional information beyond the validity of the statement [2]. The concept of ZKPs was first introduced by Goldwasser, Micali, and Rackoff in 1985 [3], and since then, it has been extensively studied and applied in various domains, including authentication, digital signatures, and blockchain technology [16, 17].

A ZKP protocol must satisfy three main properties [18]:

1. Completeness: If the statement is true, an honest prover should be able to convince an honest verifier of its validity.

2. Soundness: If the statement is false, no prover (even a dishonest one) should be able to convince an honest verifier that it is true, except with a negligible probability.

3. Zero-knowledge: The verifier should not learn any information from the proof except for the validity of the statement.

There are several types of ZKPs, including interactive and non-interactive proofs, as well as specific constructions such as zk-SNARKs (Zero-Knowledge Succinct Non-Interactive Arguments of Knowledge) [1, 19] and zk-STARKs (Zero-Knowledge Scalable Transparent Arguments of Knowledge) [20–22]. These constructions have been used to enable privacy-preserving applications in blockchain systems, such as confidential transactions [23], anonymous voting [24], and verifiable computation [25].

### 2.2 Merkle Trees

Merkle trees, also known as hash trees, are a fundamental data structure used in many blockchain protocols to enable efficient and secure verification of large datasets [26]. A Merkle tree is a binary tree in which each leaf node contains the hash of a data block, and each non-leaf node contains the hash of its child nodes' hashes [26, 27]. The root of the tree is a single hash value that represents the entire dataset.

The main advantage of Merkle trees is that they allow for efficient verification of the integrity and inclusion of specific data elements without requiring the verifier to store or process the entire dataset. To prove the inclusion of a data element in a Merkle tree, a prover needs to provide a Merkle proof, which consists of the hashes along the path from the leaf node (representing the data element) to the root of the tree. The verifier can then reconstruct the root hash using the provided hashes and compare it with the known root hash to verify the inclusion of the data element [28].

Merkle trees have been widely adopted in blockchain systems, such as Bitcoin [29] and Ethereum [30], to store transactions, account balances, and other critical information in a tamper-evident and space-efficient manner. They play a crucial role in enabling light clients, such as mobile wallets, to securely interact with the blockchain without having to download and process the entire blockchain history.

### 2.3 Proof Aggregation Techniques

As blockchain networks scale and the number of data elements stored in Merkle trees grows, the size and complexity of Merkle proofs can become a bottleneck in terms of verification efficiency and data communication overhead [8, 28]. To address this issue, proof aggregation techniques have been proposed to combine multiple proofs into a single, compact proof [31, 32].

The most common proof aggregation approach is based on AND logic, where the aggregated proof is considered valid only if all the constituent proofs are valid [31, 32]. In the context of Merkle tree inclusion proofs, AND aggregation allows the prover to combine the proofs for multiple data elements into a single proof. However, the verifier still needs to process all the leaf hashes to validate the aggregated proof, leading to high verification complexity, especially for large Merkle trees.

Other proof aggregation techniques have been explored in the literature, such as batch verification [6, 21], recursive proof composition [1, 12], and probabilistic proof aggregation [33, 34]. These approaches aim to improve the efficiency and scalability of proof systems by reducing the verification complexity and data communication overhead. However, they often come with trade-offs in terms of proof size, security guarantees, and universality.

In this paper, we propose a novel proof aggregation approach based on OR logic, which enables the generation of compact and universally verifiable proofs for Merkle tree inclusion. Our scheme leverages the properties of OR logic to create a single, aggregated proof that can be verified using any valid leaf hash, thus significantly reducing the verification complexity and data communication overhead compared to AND aggregation. The following sections will introduce our proposed scheme and provide a comparative analysis with existing proof aggregation techniques.

### 3. Enhanced Aggregation Logic

In this section, we introduce a novel approach to proof aggregation in zero-knowledge proof systems, which addresses the limitations of traditional AND aggregation logic. We propose an enhanced aggregation scheme based on OR logic that enables the generation of compact and universally verifiable zk-proofs for Merkle tree inclusion. The proposed scheme offers improved scalability, flexibility, and efficiency compared to existing approaches, making it a promising solution for large-scale Merkle tree implementations in blockchain systems.

### 3.1. Motivation for an improved universal proof

Let $\mathcal{M}$ be a Merkle tree with $n$ leaves, where $n = 2^d$ for some integer $d \geq 0$. Each leaf is associated with a data block $b_i$, $(i = 1, \ldots, n)$ and the corresponding leaf hash is computed as $h_i = H(b_i)$, where $H(\cdot)$ is a cryptographic hash function. The Merkle tree is constructed by recursively hashing pairs of adjacent nodes until a single root hash $h_{\text{root}}$ is obtained.

In a traditional AND aggregation scheme, an aggregated zk-proof $\pi_{\text{AND}}$ is considered valid only if all the constituent zk-proofs $\pi_1, \ldots, \pi_m$ are valid. Formally, let $\pi_i$ be a proof for the validity of a leaf hash $h_i$, and let $\mathcal{V}(\pi_i, h_i)$ denote the verification function that outputs 1 if $\pi_i$ is a valid proof for $h_i$ and 0 otherwise. Then, the AND aggregation of proofs $\pi_1, \ldots, \pi_m$ is defined as (Fig. 1):

$$\pi_{\text{AND}} = \text{AND}(\pi_1, \ldots, \pi_m) \Leftrightarrow \mathcal{V}(\pi_{\text{AND}}, (h_1, \ldots, h_m)) \equiv \bigwedge_{j=1}^{m} \mathcal{V}(\pi_j, h_j).$$

In other words, the aggregated proof $\pi_{\text{AND}}$ is valid if and only if all the constituent proofs $\pi_1, \ldots, \pi_m$ are valid.

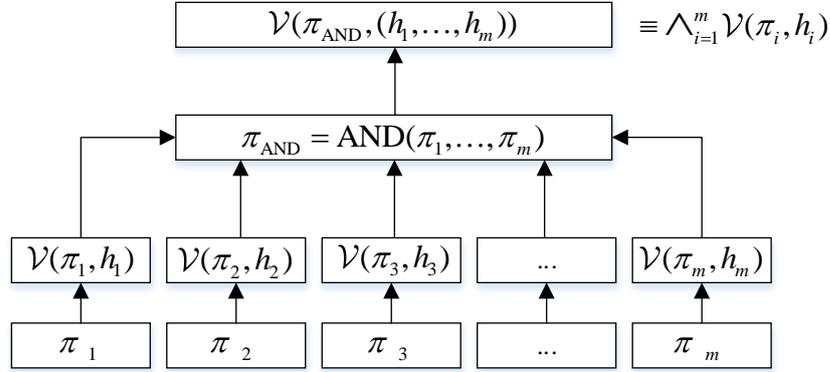

Figure 1: Aggregation logic "AND" of zero-knowledge proofs

While AND aggregation has been used effectively in various scenarios, it poses significant challenges when applied to large Merkle trees. To illustrate this, consider the problem of proving the inclusion of a single leaf $b_i$ in a Merkle tree $\mathcal{M}$. In a standard Merkle proof, the prover provides the verifier with a path of hashes from the leaf $b_i$ to the root $h_{\text{root}}$, along with the corresponding sibling hashes at each level. The verifier can then recompute the root hash and compare it with the known value to verify the inclusion of $b_i$.

However, if we were to use AND aggregation to create a single zk-proof for the inclusion of $l_i$, the prover would need to provide proofs for all the leaves in the tree, i.e., $\pi_1, \ldots, \pi_n$, where $n = 2^d$. The aggregated proof $\pi_{\text{AND}}$ would then be validated by verifying each constituent proof (Fig. 2):

$$\mathcal{V}(\pi_{\text{AND}}, (h_1, \ldots, h_n)) \equiv \bigwedge_{j=1}^{n} \mathcal{V}(\pi_j, h_j).$$

The main challenge with using AND aggregation for Merkle tree inclusion proofs is the verification complexity. While the size of the aggregated proof $\pi_{\text{AND}}$ itself may be compact, the verifier would need to be provided with all the leaf hashes $h_1, \ldots, h_n$ to validate the proof (Fig. 2). In a tree with $2^{30}$ leaves (corresponding to a 1 GB data block), this would require the prover to send and the verifier to process $2^{30}$ hash values, each typically 256 bits long, resulting in a total communication overhead of 32 GB. This makes the verification process impractical for large Merkle trees.

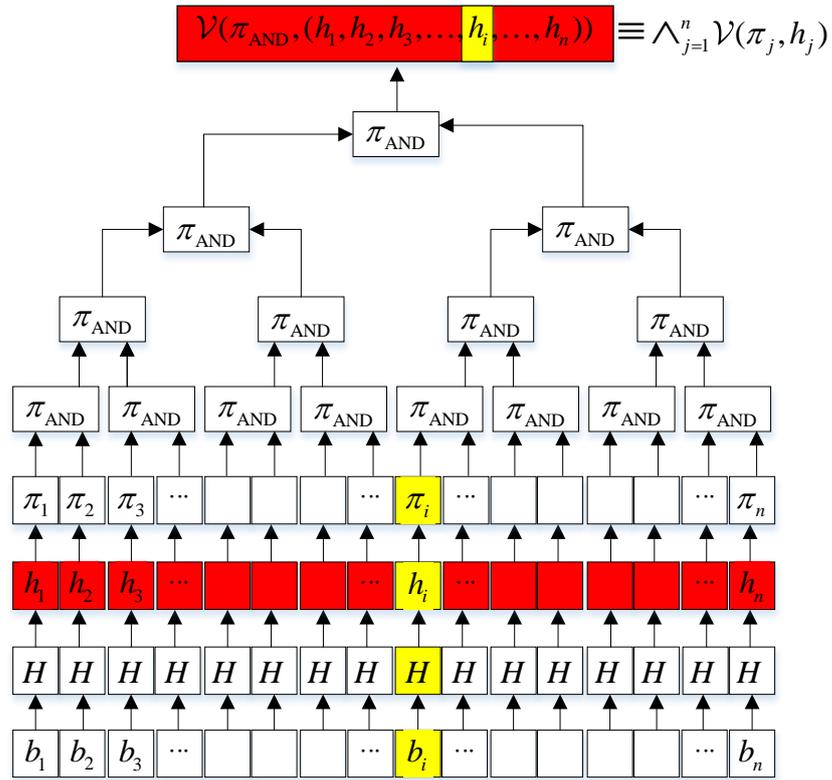

Fig. 2: AND logic to create a single zk-proof of inclusion

One way to mitigate this issue is to embed the specific Merkle path elements for a particular leaf into the final proof, as done in the Maru project [35]. This approach eliminates the need to provide all the leaf hashes during verification. However, the resulting proof is no longer universal, as it is tailored to prove the inclusion of a single, specific leaf. If the prover wants to demonstrate the inclusion of a different leaf, a new proof must be generated, embedding the corresponding Merkle path elements.

Formally, let $\pi_{AND}(b_i)$ denote the AND-aggregated proof for the inclusion of leaf $b_i$, with the Merkle path elements $h_1, h_2, ..., h_d$ for $b_i$ embedded in the proof. The verification of $\pi_{AND}(b_i)$ would only require the leaf hash $h_i$ and the root hash $h_{root}$ (Fig. 3):

$$\mathcal{V}(\pi_{AND}(b_i), (h_i, h_{root})) = 1 \Leftrightarrow b_i \in \mathcal{M}.$$

Figure 3 shows:

$x$ - public statement;

$w$ - secret witness;

$\|$ - concatenation function (combining vectors).

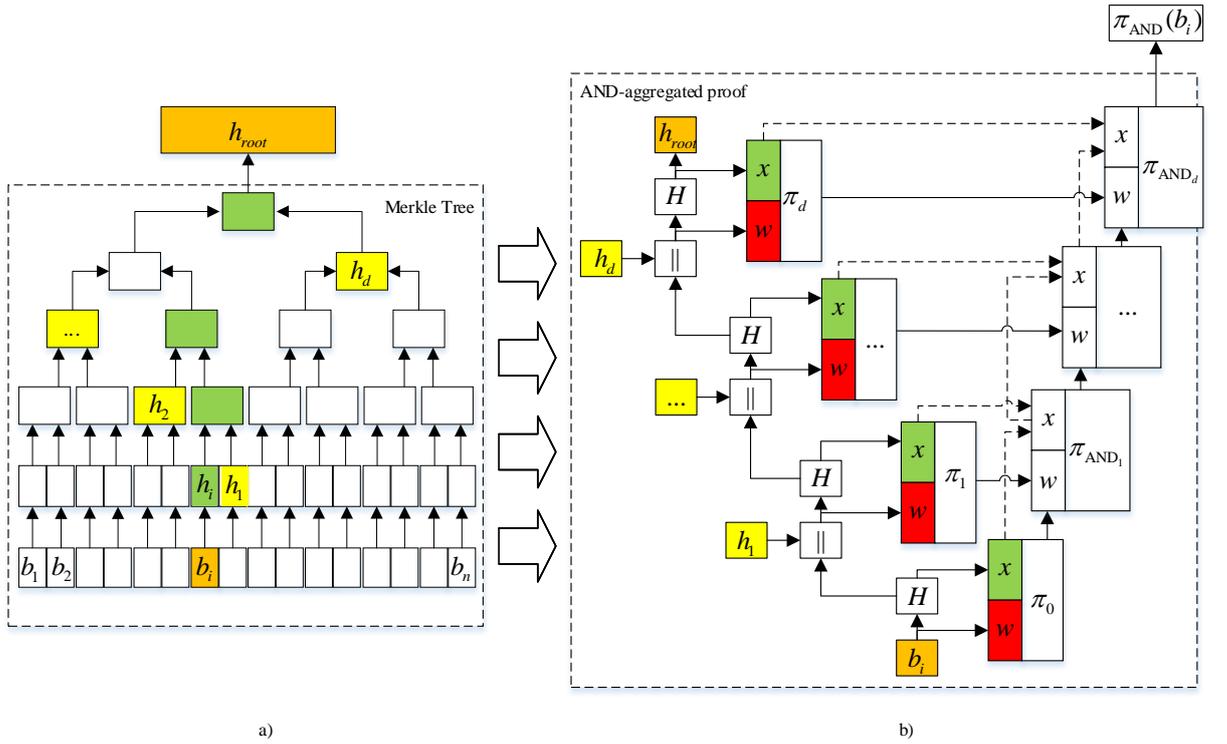

Fig. 3: Logic for generating a single inclusion proof with Merkle path embedding (as used in the Maru project [35]):

- Scheme for using the Merkle proof $h_1, h_2, ..., h_d$
- Schematic of generating the chain of proofs of computational integrity $\pi_0, \pi_1, ..., \pi_d$ and the chain of recursive proofs $\pi_{AND_1}, \pi_{AND_2}, ..., \pi_{AND_d}$ in Merkle Tree

While this approach reduces the communication overhead and verification complexity, it comes at the cost of proof universality. The prover must generate a separate proof $\pi_{AND}(b_i)$ for each leaf $b_i$, $i = 1, 2, ..., n$ they want to prove inclusion for, which can be inefficient when dealing with a large number $n$ of leaves.

In the following sections, we introduce an enhanced aggregation scheme based on OR logic that aims to address these limitations. The proposed scheme allows for the generation of a compact and universal proof of Merkle tree inclusion, which can be verified efficiently without requiring the prover to send all the leaf hashes or embed specific Merkle path elements for each leaf.

### 3.2. OR aggregation logic

To address the limitations of AND aggregation, we propose a novel proof aggregation scheme based on OR logic. In contrast to AND aggregation, OR aggregation allows for the construction of a valid proof **if at least one of the constituent proofs is valid**.

Formally, let $\pi_1, ..., \pi_m$ be proofs for the validity of leaf hashes $h_1, ..., h_m$, respectively. The OR aggregation of these proofs, denoted by $\pi_{OR}$, is defined as (Fig. 4):

$$\pi_{OR} = OR(\pi_1, ..., \pi_m) \Leftrightarrow \mathcal{V}(\pi_{OR}, h_i) \equiv \bigvee_{j=1}^{m} \mathcal{V}(\pi_j, h_j).$$

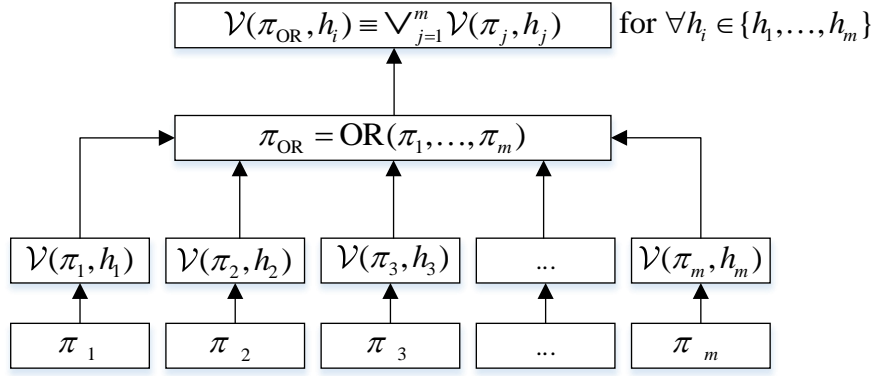

Figure 4: Aggregation logic "OR" of zero-knowledge proofs

In other words, the aggregated proof $\pi_{OR}$ is valid if and only if at least one of the constituent proofs $\pi_1,\ldots,\pi_m$ is valid. This means that if at least one valid value $h_i \in \{h_1,\ldots,h_m\}$ is supplied to the input of the proof checking function $\mathcal{V}(\pi_{OR}, h_i)$, we will receive confirmation of the aggregate proof of inclusion

$$\mathcal{V}(\pi_{OR}, h_i) \equiv \vee_{j=1}^m \mathcal{V}(\pi_j, h_j) = 1, \quad \text{for } h_i \in \{h_1,\ldots,h_m\}.$$

The OR aggregation logic enables a more efficient traversal of the Merkle tree, where proofs for individual leaves can be aggregated in a way that naturally follows the tree structure.

Let $\mathcal{M}$ be a Merkle tree with $n$ leaves, and let $b_1,\ldots,b_n$ be the leaf nodes with corresponding hashes $h_1,\ldots,h_n$. The OR aggregation process begins at the leaves and progressively combines the proofs of adjacent nodes to form proofs for their parent nodes.

At the first level of aggregation, the proofs for adjacent leaf nodes are combined using OR logic:

$$\pi_{OR} = OR(\pi_{left}, \pi_{right}),,$$

where $\pi_{OR}$ is the aggregated proof for the parent node of leaves $b_{left}$ and $b_{right}$. This process is repeated iteratively for each subsequent level of the tree, combining the proofs of sibling nodes to form a proof for their parent node (Fig. 4).

The aggregation process continues until a single proof $\pi_{OR\,root}$ is obtained for the root of the tree. This proof has the property that it can be validated by providing any one of the valid leaf hashes as input:

$$\mathcal{V}(\pi_{OR\,root}, h_i) = 1 \Leftrightarrow b_i \in \mathcal{M}, \quad \text{for } i = 1,\ldots,n.$$

The OR aggregation scheme allows for the generation of a compact and universally verifiable proof of Merkle tree inclusion, as will be discussed in the following section.

### 3.3. Generating a universal proof for Merkle tree inclusion

The OR aggregation scheme described in the previous section enables the generation of a universal proof that succinctly attests to the inclusion of any valid leaf in the Merkle tree. The process of generating this universal proof, denoted by $\pi_{\text{OR root}}$, consists of the following steps (Fig. 5):

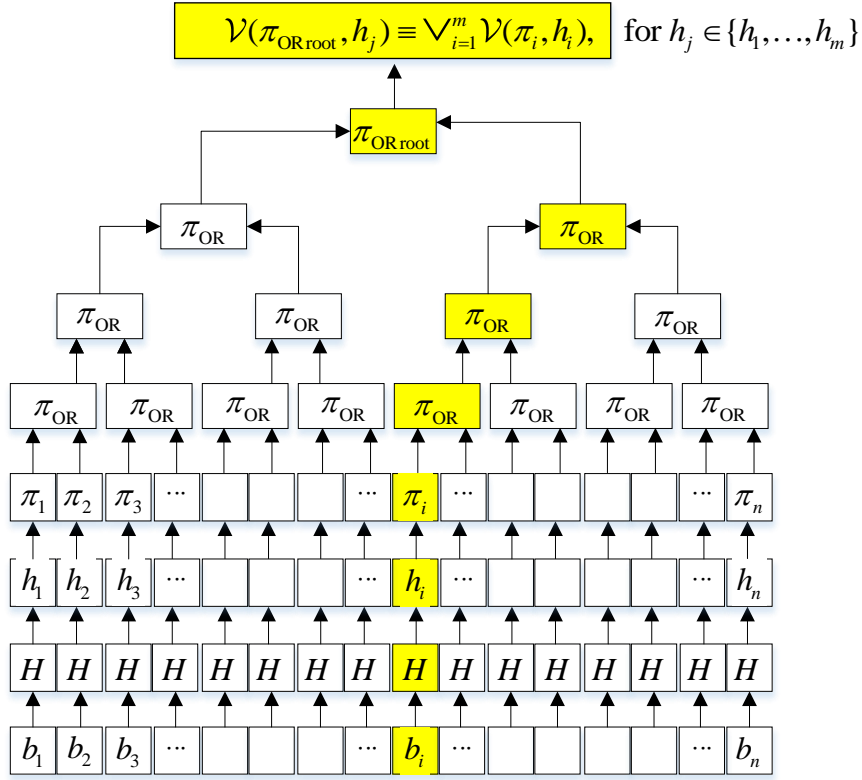

Fig. 5: OR logic to create a single zk-proof of inclusion

1. Generate proofs for each leaf: For each leaf node $b_i$ $(i = 1,\ldots,n)$ in the Merkle tree, generate a zero-knowledge proof $\pi_i$ that attests to the correctness of the leaf hash $h_i$. This can be done using a suitable zero-knowledge proof system, such as zk-SNARKs or zk-STARKs.

2. Aggregate proofs using OR logic: Starting from the leaves, recursively aggregate the proofs of adjacent nodes using OR logic, as described in Section 4.2. At each level, the proofs of sibling nodes are combined to form a proof for their parent node. This process is repeated until a single proof $\pi_{\text{OR root}}$ is obtained for the root of the tree.

3. Output the universal proof: The aggregated proof for the root of the Merkle tree, $\pi_{\text{OR root}}$, serves as the universal proof of inclusion. This proof has the property that it can be validated by providing any one of the valid leaf hashes as input:

$$\mathcal{V}(\pi_{\text{OR root}}, h_i) = 1 \Leftrightarrow l_i \in \mathcal{M}, \quad \text{for } i = 1,\ldots,n.$$

The resulting universal proof $\pi_{\text{OR root}}$ is compact, as its size is independent of the number of leaves in the tree. Moreover, the proof can be efficiently verified by providing any one of the valid leaf hashes, without requiring the prover to send all the leaf hashes or embed specific Merkle path elements for each leaf.

### 3.4. Advantages of the new approach

The proposed OR aggregation scheme offers several advantages over the traditional AND aggregation approach:

1. Scalability: The universal proof $\pi_{\text{ORroot}}$ generated using OR aggregation remains compact and efficiently verifiable, even for large Merkle trees. The size of the proof and the verification complexity are independent of the number of leaves in the tree, making the scheme highly scalable. In contrast, AND aggregation requires the prover to send and the verifier to process all the leaf hashes, which becomes impractical for large trees.
2. Flexibility: The universal proof can be validated using any one of the valid leaf hashes, providing flexibility in proof verification. This property is particularly useful in scenarios where the verifier may not have access to all the leaf hashes or may only be interested in verifying the inclusion of a specific leaf. In comparison, AND aggregation with embedded Merkle path elements (as used in the Maru project [35]) requires the prover to generate a separate proof for each leaf they want to prove inclusion for, which can be inefficient when dealing with a large number of leaves.
3. Efficiency: The OR aggregation scheme eliminates the need to provide all the leaf hashes during proof verification, resulting in reduced communication overhead and improved verification efficiency compared to the AND aggregation approach. The verifier only needs to process a single leaf hash to validate the universal proof, regardless of the size of the Merkle tree.
4. Compatibility: The proposed scheme is compatible with existing Merkle tree implementations and can be easily integrated into systems that already employ Merkle trees for data integrity and authentication purposes. The OR aggregation logic can be applied on top of the existing Merkle tree construction, without requiring significant modifications to the underlying data structures or cryptographic primitives.

In summary, the enhanced aggregation logic based on OR aggregation offers a promising solution to the scalability and flexibility limitations of traditional proof aggregation techniques. By enabling the generation of compact and universally verifiable proofs, this approach has the potential to significantly improve the efficiency and practicality of zero-knowledge proof systems in the context of Merkle tree inclusion proofs, particularly for large-scale blockchain applications.

## 4. Discussion and Analysis

### 4.1 Comparison of aggregation

The proposed OR aggregation logic for proof aggregation offers several advantages over the traditional AND aggregation approach. As demonstrated in the previous sections, OR aggregation enables the generation of compact and universally verifiable proofs for Merkle tree inclusion, which significantly improves the efficiency and scalability of the overall proof system.

In the case of Merkle tree inclusion proofs, the OR aggregation scheme allows for the generation of a single, compact proof that attests to the inclusion of any valid leaf in the tree. This proof can be verified by providing just one valid leaf hash, which aligns perfectly with the logic of classical Merkle proofs. In contrast, the traditional AND aggregation approach requires the prover to send and the verifier to process all the leaf hashes, leading to high communication overhead and computational complexity, especially for large Merkle trees.

Table 1 presents a comparative analysis of our proposed OR aggregation approach with the traditional AND aggregation logic and the methods used in the Maru project.

As shown in the table, our proposed OR aggregation approach achieves compact proof sizes, similar to the AND aggregation and Maru project methods. However, the key advantage of OR aggregation lies in the verification process. While AND aggregation requires the verifier to process all the leaf hashes, leading to high data communication overhead, OR aggregation enables verification with just a single leaf hash. This significantly reduces the amount of data required for proof verification, making the scheme more efficient and scalable.

Table 1: Comparative analysis of aggregation

| Approach | Proof Size | Verification Data | Universality |
|---|---|---|---|
| AND Aggregation | Compact | All leaf hashes | Universal |
| Maru Project (Embedded) | Compact | Single leaf hash | Leaf-specific |
| OR Aggregation (Proposed) | Compact | Single leaf hash | Universal |

Compared to the Maru project [35], which embeds Merkle path elements for a specific leaf into the proof, our OR aggregation approach generates a universal proof that can be verified using any valid leaf hash. This universality property is crucial, as it allows the prover to generate a single proof that can be used to attest to the inclusion of any leaf in the Merkle tree, without the need to create separate proofs for each leaf. This flexibility is particularly advantageous in scenarios where the prover needs to prove the inclusion of multiple leaves or where the verifier is interested in verifying the inclusion of a specific leaf without requiring the prover to generate a tailored proof.

**4.2 Extending the technique to new applications**

The introduction of OR aggregation logic alongside the traditional AND aggregation opens up new possibilities for constructing complex acceptance functions at the proof generation level. By combining these two aggregation functions, we can create sophisticated proof systems that cater to various business logic requirements in blockchain applications.

For instance, consider a scenario where a smart contract needs to verify that a certain condition is met by at least one participant from a group. In this case, the participants can generate individual proofs attesting to their fulfillment of the condition, and these proofs can be aggregated using OR logic. The smart contract can then verify the aggregated proof, ensuring that at least one participant has met the required condition, without needing to verify each proof separately.

Similarly, AND aggregation can be used in situations where a smart contract requires all participants to satisfy a specific condition. The individual proofs can be aggregated using AND logic, and the smart contract can verify the aggregated proof to ensure that all participants have met the requirement.

Furthermore, by nesting AND and OR aggregations, we can create even more complex acceptance functions. For example, a smart contract may require that either all participants from group A OR at least one participant from group B satisfy a condition. In this case, the proofs from group A can be aggregated using AND logic, while the proofs from group B can be aggregated using OR logic. The resulting aggregated proofs can then be combined using OR logic to create a single proof that represents the complex acceptance function.

This flexibility in constructing acceptance functions at the proof level can greatly enhance the expressiveness and efficiency of blockchain applications. Instead of relying on smart contracts to verify each condition separately, the verification complexity can be shifted to the proof generation phase, allowing for more streamlined and cost-effective contract execution.

Moreover, the ability to create complex acceptance functions using AND and OR aggregations can enable the development of new types of blockchain applications that were previously infeasible or impractical due to the limitations of traditional proof systems. For instance, applications involving multi-party computation, confidential transactions, or complex voting schemes can benefit from the expressiveness and efficiency provided by the combination of AND and OR aggregations.

## 5. Conclusion

In this paper, we have introduced a novel proof aggregation technique based on OR logic, which addresses the limitations of traditional AND aggregation in the context of Merkle tree inclusion proofs. The proposed OR aggregation scheme enables the generation of compact and universally verifiable proofs, allowing for efficient and scalable verification of Merkle tree inclusion.

We have formally defined the OR aggregation logic and described the process of generating a universal proof for Merkle tree inclusion using this approach. The resulting proof is compact in size and can be verified using any single valid leaf hash, providing a significant advantage over the traditional AND aggregation method, which requires the verifier to process all leaf hashes.

Through a comparative analysis, we have demonstrated the benefits of our proposed approach in terms of proof size, verification data, and universality. The OR aggregation scheme achieves compact proofs, requires minimal verification data, and generates a universal proof that can be used to attest to the inclusion of any valid leaf in the Merkle tree.

Furthermore, we have discussed the potential of combining OR and AND aggregation logics to create complex acceptance functions at the proof generation level. This flexibility enables the development of expressive and efficient proof systems that can cater to various business logic requirements in blockchain applications.

The proposed techniques have the potential to significantly enhance the scalability, efficiency, and expressiveness of zero-knowledge proof systems in the context of Merkle tree inclusion proofs and beyond. As the adoption of zero-knowledge proofs continues to grow in blockchain applications, the ability to construct flexible and efficient proof aggregation schemes will be crucial in enabling the development of scalable and practical solutions.

Future research directions include the implementation and evaluation of the proposed OR aggregation scheme in real-world blockchain systems, the exploration of hybrid aggregation techniques that combine OR and AND logics, and the investigation of other aggregation logics that can further enhance the expressiveness and efficiency of proof systems.